\documentclass[conference,10.5pt]{IEEEtran}
\IEEEoverridecommandlockouts
\usepackage{cite}
\usepackage{amsmath,amssymb,amsfonts}
\usepackage{algorithm}% http://ctan.org/pkg/algorithms
\usepackage{algcompatible}% http://ctan.org/pkg/algorithmicx

\algnewcommand\algorithmicreturn{\textbf{return}}
\algnewcommand\RETURN{\State \algorithmicreturn}%
\usepackage{graphicx}
\usepackage{color}
\usepackage{subfig}
\usepackage{stackengine}
\usepackage{comment}
\usepackage[normalem]{ulem}
\usepackage{tabularx}
\usepackage{bm}
\usepackage{bbold}
\usepackage{mathtools, nccmath}
\usepackage{multirow}
\usepackage{stfloats}

\usepackage[singlespacing]{setspace} 

\usepackage{lettrine}
\usepackage[font=scriptsize]{caption}

\usepackage[colorlinks, citecolor=blue]{hyperref} % to highlight references with colors

\usepackage[flushleft]{threeparttable} % http://ctan.org/pkg/threeparttable
\usepackage{booktabs}

\usepackage{textcomp}
\usepackage{xcolor}

\def\BibTeX{{\rm B\kern-.05em{\sc i\kern-.025em b}\kern-.08em
    T\kern-.1667em\lower.7ex\hbox{E}\kern-.125emX}}
    
\DeclareMathOperator*{\argmax}{argmax}

\usepackage{orcidlink}

\usepackage{xspace}
\usepackage{ifthen}
\usepackage[inline]{enumitem}

\newboolean{showcomments}
\setboolean{showcomments}{true}
\ifthenelse{\boolean{showcomments}}
{
}

\usepackage{subfig}
\usepackage{fancyhdr}

\newcommand*{\affmark}[1][*]{\textsuperscript{#1}}

\begin{document}

\title{\huge Extended Reality (XR) Codec Adaptation in 5G using Multi-Agent Reinforcement Learning with Attention Action Selection}

\author{\small\IEEEauthorblockN{Pedro Enrique Iturria-Rivera\orcidlink{0000-0002-8757-2639}\affmark[1], \IEEEmembership{Student Member,~IEEE}, Raimundas Gaigalas\affmark[2],  Medhat Elsayed\affmark[2], \\ Majid Bavand\affmark[2], Yigit Ozcan\affmark[2] and Melike Erol-Kantarci\orcidlink{0000-0001-6787-8457}\affmark[1], \IEEEmembership{Senior Member,~IEEE}}
\IEEEauthorblockA{\affmark[1]\textit{School of Electrical Engineering and Computer Science, University of Ottawa, Ottawa, Canada}\\\affmark[2]\textit{Ericsson Inc., Ottawa, Canada}}Emails:\{pitur008, melike.erolkantarci\}@uottawa.ca, \{raimundas.gaigalas, medhat.elsayed,\\ majid.bavand, yigit.ozcan\}@ericsson.com\vspace{-1em} }

\maketitle
\begin{abstract}
\textbf{Extended Reality (XR) services will revolutionize applications over 5$^{th}$ and 6$^{th}$ generation wireless networks by providing seamless virtual and augmented reality experiences. These applications impose significant challenges on network infrastructure, which can be addressed by machine learning algorithms due to their adaptability. This paper presents a Multi-Agent Reinforcement Learning (MARL) solution for optimizing codec parameters of XR traffic, comparing it to the Adjust Packet Size (APS) algorithm. Our cooperative multi-agent system uses an Optimistic Mixture of Q-Values (oQMIX) approach for handling Cloud Gaming (CG), Augmented Reality (AR), and Virtual Reality (VR) traffic. Enhancements include an attention mechanism and slate-Markov Decision Process (MDP) for improved action selection. Simulations show our solution outperforms APS with average gains of $30.1 \%$, $15.6 \%$,  $16.5 \%$ $50.3 \%$ in XR index, jitter, delay, and Packet Loss Ratio (PLR), respectively. APS tends to increase throughput but also packet losses, whereas oQMIX reduces PLR, delay, and jitter while maintaining goodput.
}
\end{abstract}

\vspace{-5mm}
\section{Introduction}

\lettrine[findent=1pt]{\textbf{T}}{}he sixth generation (6G) wireless systems is anticipated to drive several new Internet of Everything (IoE) applications. 6G will be identified mainly by two fundamental characteristics: self-sustainability and proactiveness \cite{Khan2022}. The previous terms refer to the capability of 6G to perform efficient adaptability and fulfillment of the extreme requirements of the emerging IoE services. Cloud Gaming (CG) and Extended Reality (XR), including Virtual Reality (VR), Augmented Reality (AR), and Mixed Reality (MR) are considered new emergent applications in 6G. XR technology has been around for some time, however, its impact and potential applications have been extended in 6G from the well-known gaming and smartphone applications to the healthcare industry and other verticals \cite{Ahmad2023}. For this reason, 3GPP and others have shown their interest in the standardization of XR in 5G since 2016. More specifically, 3GPP release-16 studies VR Quality of Experience (QoE) relevant metrics for user experience \cite{3gpp1}. Furthermore, 3GPP presents, in \cite{3gpp2} an introductory report in relation to XR in 5G where 23 possible use cases are presented, altogether with a standardized 5G QoS Identifier (5QI) mapping to Quality of Service (QoS). 
\begin{figure}[t]
\center
  \includegraphics[scale=0.17]{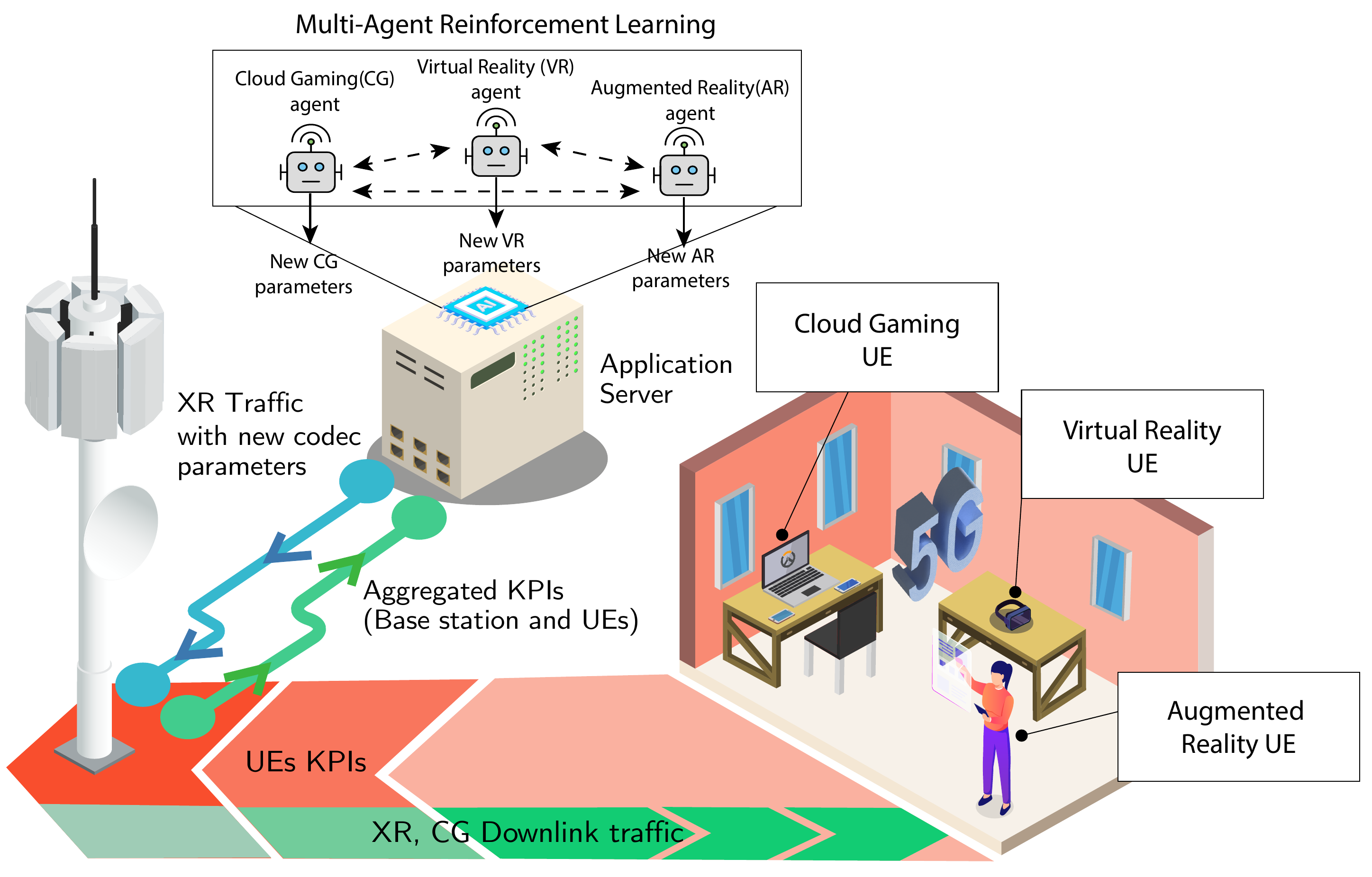}
  \setlength{\belowcaptionskip}{-5pt}
  \caption{An AI-powered application server exchanges aggregated KPIs from the BS and UEs to decide on the proper XR and CG codec parameters to satisfy XR/CG QoE requirements.} 
  \label{system_overview1}\vspace{-1em}
\end{figure} 
In this work, as indicated in Fig. \ref{system_overview1}, we utilize a 5G network with User Equipments (UEs) receiving diverse XR traffic (AR, VR) and CG in the downlink direction. The users report the Key Performance Indicator (KPI) metrics to the base station (BS) to be further aggregated with BS's local KPIs and shared with an application server. Afterward, the application server decides the new optimized codec parameters for the future XR traffic. The codec parameters consist of the data rate or the frame per second (FPS) of the XR traffic. Instead of modeling this problem in a centralized architecture, we propose the usage of a more realistic but challenging Multi-Agent Reinforcement Learning (MARL) to optimize the codec selection using cross-layer information. More specifically, we model this problem as a multi-agent scenario where three agents (AR, VR, CG) will act as a team to maintain fairness and quality of experience among the XR users. We choose the QMIX (Mixture of Q-Values) algorithm over other centralized-training decentralized-execution (CTDE) methods due to its proven excellent performance. More specifically, we utilized a modified QMIX algorithm named Optimistic QMIX (oQMIX) and we leveraged an attention technique to improve the learning performance of agents. In this paper, we refer to QMIX and oQMIX, assuming that they incorporate our proposed attention mechanism. This is not to be mistaken with the original algorithms that lack this feature.

We compare our attention-based QMIX and oQMIX algorithms with a state-of-the-art XR loopback mechanism called Adjust Packet Size (APS) \cite{Bojovic2023}. Our results show average gains of oQMIX and QMIX over APS of $30.1 \%$, $15.6 \%$,  $16.5 \%$ $50.3 \%$, and $17.6 \%$, $13.2 \%$,  $11.2 \%$, $7.86 \%$ with respect XR index, jitter, delay, and Packet Loss Ratio (PLR), respectively.

The rest of this paper is organized as follows. Section \ref{Section2} presents the existing works related to XR traffic codec and video streaming optimization. In Section \ref{section3} a brief system model is described. Section \ref{section4} introduces some background on QoE in XR and presents the MDP preliminaries of the proposed algorithm and attention mechanism. Additionally, section \ref{section5} presents the simulation results of our proposed scheme as well as the presented baseline. Finally, Section \ref{Section6} concludes the paper.

\section{Related work}\label{Section2}
In recent years, XR traffic has attracted the interest of academia and industry due to its stringent requirements that have posed great challenges to 5G and beyond networks. XR, together with Cloud Gaming (CG) is currently one of the most important 5G media applications under consideration in the industry \cite{3gpp3}.  XR is focused on creating immersive experiences that blend the digital and physical worlds. Cloud gaming, on the other hand, is specifically about delivering gaming content via the cloud, reducing the need for powerful hardware on the user's end. Recently, some studies have proposed adaptive mechanisms to improve KPIs for XR. For instance in the works \cite{Lagen2023, Bojovic2023}, the authors study novel Quality of Service (QoS) control procedures to handle the peculiarities of XR traffic. In addition, it presents three traffic adaptation mechanisms at the application layer (named XR loopback) to improve performance between XR applications and the 5G Radio Access Network (RAN). Each mechanism consists either of adjusting in real-time XR codec parameters such as data rate or FPS or adjusting both of them concurrently. In conjunction with the previous mechanism, the authors propose a QoS-based scheduler and conclude that even though the scheduler improves the classical resource allocation schedulers such as Proportional Fairness (PF) the loopback mechanisms contribute the most to improving QoS metrics. However, some unwanted behavior can be seen in the proposed adaptive mechanisms such as unfairness among flows and increasing PLR when the channel conditions worsen.

On the other hand, several works in the literature focus their interest on codec adaptation in the context of video streaming and RL. For instance, in \cite{Sengupta2018} the authors optimize video streaming bitrate using an Actor-Critic RL architecture while considering users' preferences. In addition, in \cite{Tang2021} the authors study video streaming chunk representation by using a two-layer deep-RL framework: at the physical layer adjusting beamforming parameters and at the application layer the video chunk representation. Finally \cite{Turkkan} proposes GreenABR, an energy-aware adaptive bitrate (ABR) algorithm based on deep RL. The authors in the former work utilize real cellphone terminal energy consumption data to decide on the best bitrate without jeopardizing user QoE. In the next section, we introduce the system model and the details of our proposal.   

\section {System Model} \label{section3}

In this work, we utilize a 5G network with  $N$ XR users receiving downlink traffic from a BS. The XR traffic is generated from an application server that receives every $t_w$ window of time, feedback from the BS. The BS aggregates its own and users' KPIs and exchanges it with the application server. Upon the new feedback, the application server decides the future XR codec parameters to maintain QoE metrics.  Our scenario consists of 3 different XR users attached to one BS. To simulate a loaded network, some parameters of the simulation are adjusted such as the transmission Radio Link Control buffer's (Unacknowledged Mode) capacity and the available bandwidth. Furthermore,  UEs are located within three rings as in Fig. \ref{system_overview} (a). Finally, the modeled XR traffic complies with the 3GPP release-17 study \cite{3gpp3}.

\section{XR Codec Adaptation-based Multi-Agent Reinforcement Learning } \label{section4}
In this section, we introduce the details and considerations in the design of the XR Codec Adaptation Multi-Agent Reinforcement Learning algorithm. We begin by introducing some background on QoE in the context of XR and follow with the specifics of the QMIX algorithm and attention action selection. 

\subsection{Quality of Experience in XR}\label{XR_section}
Quality of Experience (QoE) is a well-known metric in multimedia-related topics. It is subjective to the observer and thus, difficult to measure objectively. For instance, as mentioned in \cite{Brunnstrom2013}, QoE can be defined as ``the degree of delight or annoyance of the user of an application or service. 
The Mean Opinion Score (MOS) is one of the methods utilized in the literature to measure QoE. This metric is calculated by showing different video streams to a group of observers. The opinions are classified into a rank of five levels where 5 indicates an excellent quality (imperceptible impairment) and 1 bad quality (very annoying). Despite the subjectiveness of the metric, some efforts have been made to quantify QoE with Key Performance Metrics at the Radio Access Network (RAN) level. For instance, in \cite{3gpp2} 3GPP identifies the QoS requirements for different XR services. Furthermore, according to the agreements in 3GPP RAN1 meeting \cite{3gpp1}, some combinations of Packet Success Ratio (PSR) and Packet Delay Budget (PDB) should be evaluated to measure QoE in VR/AR traffic. Based on the previous proposals, in \cite{Dou2021} the authors propose a coarse-grained XR Quality Index (XQI) that consists of a mapping between a subjective metric such as MOS and 3GPP's latest documentation. 
In the following subsection, we introduce the preliminaries of the attention-based CTDE algorithm oQMIX.

\subsection{QMIX: Monotonic Value Function Factorisation}
Numerous MARL algorithms can be found throughout the existing literature that utilize the  CTDE paradigm. Differently from Independent Learning (IL) --which does not consider cooperation among agents-- and  Centralized Learning (CL) --which assumes complete access to all agents' information--, CTDE algorithms balance both sides of the spectrum. Value Decomposition Networks (VDN) and QMIX\cite{Rashid2018} are two CTDE state-of-the-art algorithms. Both perform value function factorization among the agents by considering the assumption that the multi-agent system joint action-value function can be decomposed into individual agent’s value functions. Meanwhile, VDN has proven to be quite successful in many RL tasks and some recently in wireless access networks \cite{10293832}, QMIX has demonstrated its capacity to propose richer action-value functions without full factorization of decentralized policies. 

\begin{figure*}[t]
\center
  \includegraphics[scale=0.23]{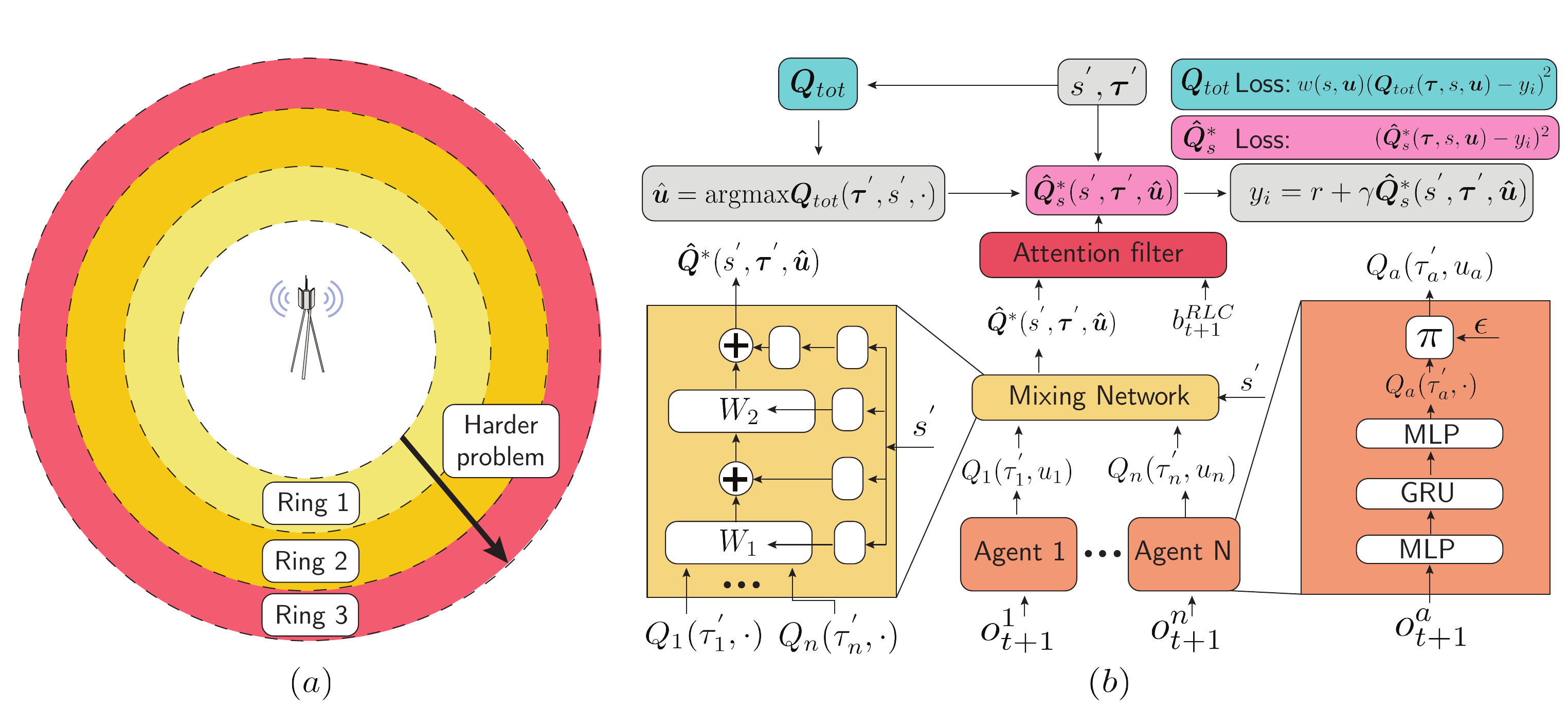}
  \setlength{\belowcaptionskip}{-5pt}
  \caption{$\bm{(a)}$ Illustration of user locations in three coverage regions. When users are located in outer rings the solution becomes harder due to the reduced action set that satisfies QoE requirements. $\bm{(b)}$ Overview of the oQMIX algorithm with action prohibition. } 
  \label{system_overview}
  \vspace{-3mm}
\end{figure*} 

\subsection{Optimistic Weighted QMIX}\label{optimistic}

In \cite{Rashid2020}, the authors introduce two distinct weighting schemes for the QMIX algorithm. QMIX, much like VDN, operates as a fully cooperative algorithm, involving a factorization of the value function among the participating agents. Instead of employing an additive approach for mixing strategies, QMIX leverages hypernetworks to enforce a monotonic behavior in its Q-function. However, a limitation of QMIX lies in its presumption of equal importance across actions, which can result in suboptimal policy outcomes. To address this shortcoming, the authors present the Optimistic Weighted QMIX, which allows for the assignment of individual weights to each Q-function, thereby facilitating better collective action decisions. The weighting parameter $w: S \times \bm{U} \rightarrow (0, 1]$ is defined as follows:

\begin{align} 
     w(s,\bm{u}) =\begin{cases} 
                            1  &  Q_{tot}( \bm{\tau},s,\bm{u}; \theta') < y_i,  \\
                               \alpha &  otherwise \\                   
                            \end{cases}
                            \label{opw}
\end{align}

where $y_i = r_i + \gamma Q_{tot}(s,o; \theta)$. Additionally,  $\bm{\tau}$, $\bm{u}$,  $s$, $o$,  $\theta$,  $\theta'$ and $\alpha \in (0,1]$ corresponds to the action-observation history, the agents' joint action, the observation state, the single agent's action, the target policy, the evaluation policy, and a predefined weight, respectively. As observed in Eq. \ref{opw} the contribution of actions considered suboptimal is reduced.

\subsection{Attention action selection and Slate-Markov Decision Process in QMIX}\label{attention-mech} \vspace{-0.5mm}
Action space size is one of the main challenges at the exploration and exploitation stages in MARL. This well-known issue is more noticeable in fully-centralized methods, yet it affects greatly CTDE methods as well. Some techniques have been discussed in the literature to reduce such dimensionality. Among them, attention mechanisms offer the capacity to give different levels of importance to different parts of the state space when deciding on the best action to take \cite{Vaswani2017}. In addition, we model our problem as a special type of Markov Decision Process (MDP) called Slate-MDP \cite{SunehagEDZVC15}. Such MDP allows to selection slate or a set of actions simultaneously rather than selecting a single action at a time. The combination of previous techniques allows us to formalize our proposal as:  

\textbf{Definition 1}. Let $\mathcal{M}=\left\langle S,U,P,R,Z,O,\gamma\right\rangle$ be an MDP that defines a fully cooperative Decentralized Partially Observable Markov Decision Process (Dec-POMDP). Let $\kappa : \mathcal{S} \times U^l \rightarrow U$. At each environment step, each agent $n \in N$ chooses an action $u^n\in U^l$, forming a joint action $\mathbf{u}^l\in\mathbf{U}^l$. $s \in S$ describes the state of the environment. After selecting all actions, this causes a transition on the environment as $P'(s'|s,\mathbf{u}^l):S\times\mathbf{U}^l\times S\rightarrow [0,1]$. The team reward is calculated as  $R'(s,\mathbf{u}^l):S\times\mathbf{U}^l\rightarrow\mathbb{R}$ and $\gamma\in[0,1)$ is a discount factor. Each agent draws individual observations $z\in Z$ according to observation function $O(s,a):S\times N\rightarrow Z$. Now, the tuple  $\left\langle S,U^l,P',R',Z,O,\gamma\right\rangle$ is called a slate dec-POMDP with an underlying MDP $\mathcal{M}$ and action-execution $\kappa$. 

We leverage the attention mechanism by selecting a portion of the agent's observation to provide an available slate of actions. This can be seen in Fig. \ref{system_overview} (b), where we incorporate an attention filter after the mixing network of the oQMIX algorithm. More specifically, the transmission RLC buffer capacity ratio $\bm{(b)}$ is utilized to create a mapping between certain predefined buffer occupancy ranges and subsets of the action set $\mathbb{A}_{t}^n$ of the agent $n^{th}$ at each time step $t$ as follows:
\begin{align}
    f:  \mathbb{A}_{t}^n \rightarrow \bm{b}_t
\end{align}

This mapping is used in Algorithm \ref{oqmix}, line 13 by the function \textsc{Is\_Action\_Disabled}. This function looks in each time step for the set of disabled actions given the observed RLC buffer capacity. If the chosen action by the policy is in such a set, then the action selection procedure will repeat. The subsets are adaptively formed by the triggering of the done condition. The done condition is a well-common Boolean variable that informs when a game becomes unsolvable in RL. In our case, the done condition becomes true when any of the XR flows’ throughput becomes zero upon the agent’s action selection. 

\begin{algorithm}[h!]

\scriptsize 
	\caption{Optimistic Weighted QMIX with attention action selection}
	\label{oqmix}
	\begin{algorithmic}[1]
		\STATE Let $\mathbb{A}_{t}^n$ be a vector of the set of disabled actions per $n^{th}$ agent, size $|\bm{b}|_t$ in step $t$ and  $\hat{\mathbb{T}}_t$ be the vector of the observable throughput of all the XR flows at step $t$. Let $\bm{b}$ be a vector of predefined transmission RLC buffer's capacity ranges.   Initialize $\theta$ and $\theta'$ indicating the target and evaluation recurrent policies, the parameters of mixing network, agent networks, and hypernetwork. Set the learning rate $\alpha$ and replay buffer $\mathcal{D} = \left\{ \right\}$
		\STATE $\text{episode} = 0, \theta^{'} = \theta$  
       
		\FOR{environment episode $e\gets1$ \textbf{to} $E$}
		\STATE{$t = 0, o_0=\text{initial state}$}
		\WHILE{$done \neq True $ and $t < \text{steps/episode limit}$}
		\FOR{each agent $m$}
		\STATE $\tau^n_t = \tau^n_{t-1} \cup \{(o_t, u_{t-1})\}$
		\STATE $\epsilon = \text{epsilon-schedule}(\text{step})$

        \IF {$\epsilon \geq \mathcal{N}(0,1)$}
        \STATE{$randint(1, |U|^n); |U|^n \subset \mathbb{Z}^+$}
        \ENDIF

        \STATE {$u^n_t= \label{choose_actions}
		\begin{cases}
			\argmax_{u^n_t}Q(\tau^n_t, u^n_t)~~~~\text{with probability }1 - \epsilon\\
			randint(1, |U|^n); |U|^n \subset \mathbb{Z}^+ ~~ \text{with probability }\epsilon\\         
		\end{cases}$}

        \IF {\textsc{Is\_Action\_Disabled($u^n_t,\mathbb{A}_{b,s}^n$)}}
            \STATE Go to line 9.
        \ENDIF
      
		\ENDFOR
		\STATE Get reward $r_t$ and next state $s_{t+1}$
		\STATE $\mathcal{D} = \mathcal{D} \cup \left\{(s_t, \mathbf{u}_t, r_t, s_{t+1} )\right\}$
		\STATE{$\text{steps}=\text{steps}+1$}

         \IF {$\nexists x \in \mathbb{\hat{T}}_t : x = 0$}
            \STATE {$done = True$}
            \STATE Update $\mathbb{A}_{b,s}^n$; $\mathbb{A}_{b,s}^n$ =  $\mathbb{A}_{b,s}^n \cup u^n_t$ 
        \ENDIF
  
		\ENDWHILE
  
		\IF {$|\mathcal{D}| > \text{batch-size}$}
		\STATE {b $\leftarrow$ random batch of episodes from $\mathcal{{D}}$}
		\FOR {each timestep $t$ in each episode in batch $v$}
		\STATE {$Q_{tot} = \textit{Mixing-network~}(Q_1(\tau^1_t,u_t^1),\dots,Q_n(\tau^n_t,u_t^n)$;}  
           \STATE {\textit{Hypernetwork}$(s_t; \theta))$}
		\STATE {Calculate target $Q_{tot}$ using $\textit{Mixing-network}$ with $\textit{Hypernetwork}(s_t; \theta^{'}))$}
		\ENDFOR
		
  \STATE {$\mathcal{L}(\theta)=\sum\limits_{i=1}^b\left[w \left(s, \mathbf{u}\right)\left(y_i^{tot} - Q_{tot}(\boldsymbol{\tau}, \mathbf{u}, s; \theta) \right)^2\right]$}
		\STATE {$\Delta \theta = \nabla_\theta(\mathcal{L}(\theta))^2$}
		\STATE {$\theta = \theta - \alpha \Delta \theta$}
		\ENDIF
		\IF {hard update steps have passed}
		\STATE {$\theta^{'} = \theta$} 
		\ENDIF
		\ENDFOR
 \vspace{-1mm}
	\end{algorithmic}

\end{algorithm}

\subsection{State space selection}
The state space of the $n^{th}$ agent of the multi-agent system is defined as follows: 

\begin{equation}
    s_{t}^n = \{\bm{a}_{(t-1)}, b_{t}^{RLC}, p_{t}^{XR}\}
\end{equation}

where $\bm{a}_{(t-1)}$ is a vector containing the team's previous codec parameter selection, $b_{t}^{RLC}$ the RLC buffer occupancy at the BS and the $p_{t}^{XR}$ corresponds to the Packet Delivery Ratio (PDR)\footnote{The Packet Delivery Ratio corresponds to the complement of Packet Loss Ratio (PLR).} of any of the traffic flows where $XR \in \{AR, VR, CG\}$. The first term helps to alleviate partial observability via the utilization of the Gated Recurrent Unit (GRU) layer in the Q-function as observed in Fig. \ref{system_overview}. The previous definition refers to the action space of a single agent in any of the MASs of our algorithm.

\subsection{Action space selection}\label{action_space}
In this subsection, we present the proposed general action space: 

\begin{equation}
    a_{t}^{XR} = \{d_{min}^{XR}, d_{min}^{XR} + \frac{d_{max}^{XR} - d_{min}^{XR}}{K_{o}-1},..., d_{max}^{XR}\},
\end{equation}

where $d_{min}^{XR}$ and $d_{max}^{XR}$ are predefined maximum and minimum values of the codec data rate values per XR traffic, respectively. The maximum and minimum values are defined in table \ref{net_settings}.

\subsection{Reward function }
The reward function is carefully designed to satisfy QoE requirements. It is modeled as a team reward and can be defined as follows:

\begin{equation}
 r_{t} = min( \hat{\mathcal{R}}_{t}^{XR}) 
 \end{equation}

   where  $\hat{\mathcal{R}}_{t}^{XR}$ is a vector comprising the reward of all the agents forming the team, $XR \in \{AR, VR, CG\}$ corresponds to the three types of codec adaptation agents. 

The individual reward of any of the XR codec agents can be defined in time $t$ as: 
\begin{equation} \label{penalization}
   \mathcal{R}_t^{XR} =
    \begin{cases}
     r_{t}^{XQI}   & \text{if $\nexists x \in \mathbb{\hat{T}}_t : x = 0$}\\    
      -1 & \textit{otherwise}
    \end{cases}  \\
\end{equation}

\begin{figure*}[t]
\centering
\begin{tabular}{@{}c@{}}
   \includegraphics[scale=0.75]{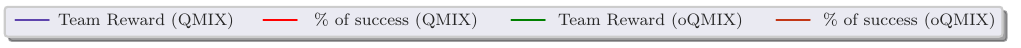}     
  \end{tabular}
\begin{tabular}{@{}c@{}}
   \includegraphics[scale=0.75]{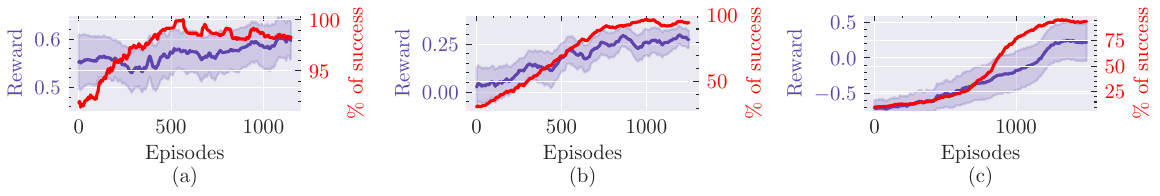} 
 \end{tabular}
  \begin{tabular}{@{}c@{}}
 \includegraphics[scale=0.75]{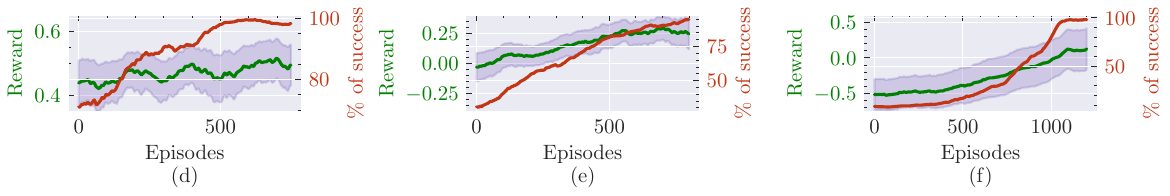} 
  
 \end{tabular}
 \caption{Convergence performance for QMIX: $\bm{(a)}$ 200 m, $\bm{(b)}$ 300 m and $\bm{(c)}$ 400 m  and oQMIX: $\bm{(d)}$ 200 m, $\bm{(e)}$ 300 m and $\bm{(f)}$ 400 m. The left y-axis and right y-axis in each subfigure indicate the team reward and $\%$ of success of each algorithm, respectively. } 
  \label{convergence} 
    \vspace{-5mm}
\end{figure*}

\begin{equation}
    r_{t}^{XQI} = 
    \begin{cases}
        1   & \text{if $p_t^{XR} \geq 99\%$ and $d_t^{XR} \leq 7$}\\
        0.75   & \text{if $p_t^{XR} \geq 99\%$ and  $7 \leq  d_t^{XR}  \leq 10$}\\
        0.5   & \text{if $p_t^{XR} \geq 95\%$ and $d_t^{XR} \leq 13$}\\
        0.25   & \text{if $p_t^{XR} \geq 95\%$ and $13 \leq  d_t^{XR}  \leq 20$}\\
        0   & \text{otherwise}\\
    \end{cases}  \\
\end{equation}
 
We present a reward that considers the XR Quality Index metric referred to in \cite{Dou2021} and section \ref{XR_section}. The previously mentioned values are summarized in Table \ref{Table1}.

\begin{table}[ht]
\scriptsize 
  \centering
  \caption{KPI Mapping for XR \cite{Dou2021}} 
  \label{Table1}
  \begin{threeparttable} 
  
  \begin{tabular}{c c }  
    \toprule
    XR Quality Index (XQI) & PDR $(\%)$, PDB (ms)\\ [1ex] % inserts table

\hline % inserts single horizontal line
5&   $(99,7)$  \\ % inserting body of the table
4 &  $(99,10)$ \\
3 &  $(95,13)$\\ 
2 &  $(95,20)$ \\
1 &  PDR $< 95$ or PDB $> 20$ \\ 
  \bottomrule
  \end{tabular}
   \end{threeparttable}
   
\end{table}
Additionally, as indicated in Eq. (\ref{penalization}), we penalize the event when, after a selected combination of codec parameters by each agent, any of the XR throughput flows (vector $\mathbb{\hat{T}}_t$) becomes $0$. Furthermore, $p_t^{XR}$ and $d_t^{XR}$ represent the average Packet Delivery Ratio (PDR) and the average delay (ms) of each XR flow during the observation window.

\subsection{Baselines: Adjust Packet Size Algorithm (APS) and QMIX}

In this work, we compare our proposed scheme with one analog XR loopback threshold-based algorithm approach described in \cite{Bojovic2023} and QMIX presented in \cite{Rashid2018}. 
\begin{algorithm}[t]

\scriptsize 
	\caption{Adjust Packet Size algorithm}
	\label{aps}
	\begin{algorithmic}[1]
		\STATE{Inputs: $p_t^{XR}$ and $a_t^{XR}$}
        \IF {$p_t^{XR} > l_{dec}^s \text{ and } p_t^{XR} < l_{dec}^q$}
            \STATE{$a_t^{XR} = \text{max}(a_{t-1}^{XR} * \alpha_{dec}^{s}, a_{min})$}
        \ELSIF{$p_t^{XR} \geq l_{dec}^q$}
            \STATE{$a_t^{XR} = \text{max}(a_{t-1}^{XR} * \alpha_{dec}^{q}, a_{min})$}
        \ELSIF{$p_t^{XR} < l_{inc}^q$}
            \STATE{$a_t^{XR} = \text{min}(a_{t-1}^{XR} * \alpha_{inc}, a_{max})$}
        \ENDIF
  
	\end{algorithmic}
\end{algorithm}

As observed in Algorithm \ref{aps}, the codec parameter $a_t^{XR}$ is adjusted by observing the PDR, $p_t^{XR}$. The configuration variables $l_{dec}^s, l_{dec}^q,l_{inc}$ correspond to PDR thresholds and  $a_{max}, a_{min}$ to the maximum and minimum value of the codec parameter $a_t^{XR}$, respectively.  In addition to the algorithmic baseline APS, we also include the results obtained by the QMIX algorithm in \cite{Rashid2018}. QMIX differently from oQMIX does not consider any Q-value function weighting strategy in the loss calculation.
\section{Performance evaluation}\label{section5}

Simulations are performed using the ns-3 New Radio (NR) module in its version 2.3 of April 2023. The previous module known also as NR-Lena is built on top of the ns-3 simulator and provides simulation for 3GPP NR non-standalone cellular networks. In addition, we use as an interface between the ns-3 and the Python-based agents the module ns-3 gym \cite{ns3gym}.

\begin{figure*}[t]
%\centering
\begin{tabular}{ll}
\includegraphics[scale=0.53]{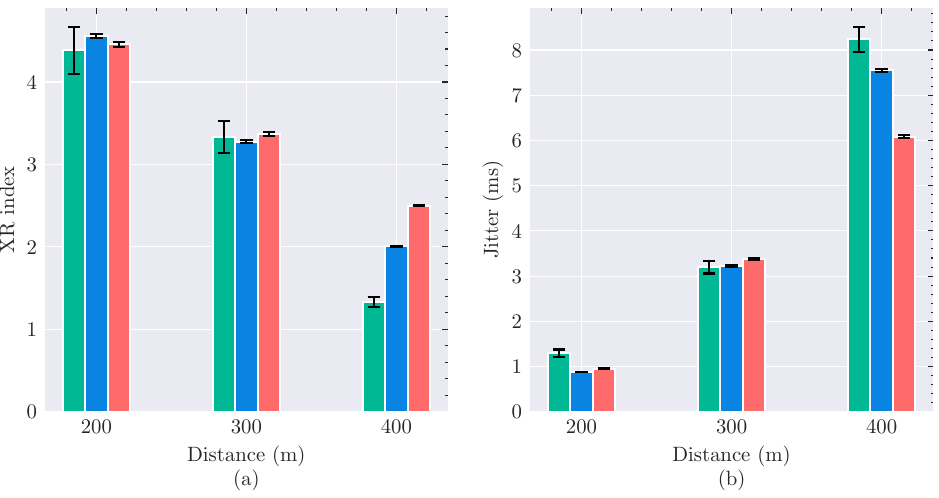}
&
\includegraphics[scale=0.53]{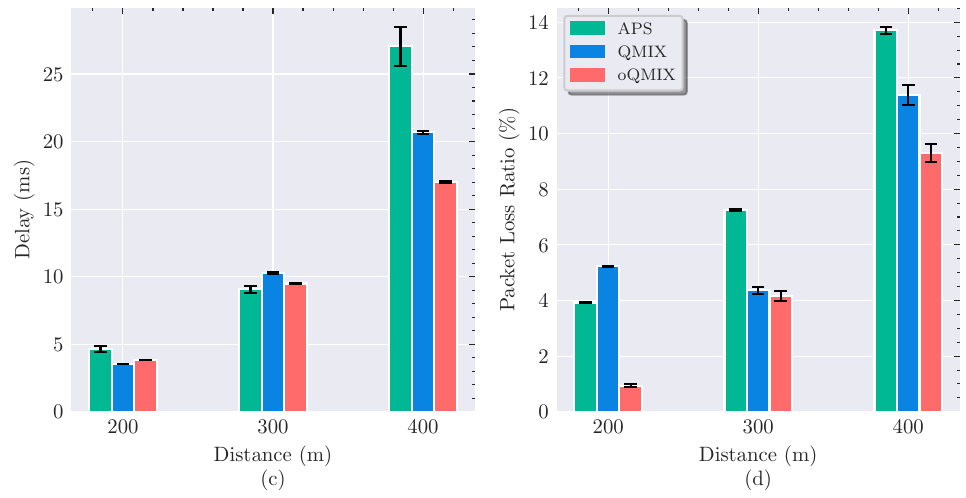}
\end{tabular}
\caption{Key Performance Indicators of interest vs. Distance $\bm{(a)}$ XR index, $\bm{(b)}$ Jitter, $\bm{(c)}$ Delay,  and  $\bm{(d)}$ Packet Loss Ratio   } 
\label{xr_jitt_del_plr_}
  \vspace{-5mm}
\end{figure*}

\subsection{Simulation Settings}
RL parameters and simulation settings are described in Table \ref{learning_settings} and \ref{net_settings}, respectively.  To study the impact of distance on the proposed algorithm and existent baseline, the users are randomly positioned in three different rings $r \in \{100-200, 200-300, 300-400\}$ m as observed in Fig. \ref{system_overview} (a). The users are randomly positioned within the inner radius of each of the rings and they maintain random mobility during the length of the simulation with a speed of $3$ m/s. The transmission RLC buffer's capacity is limited to $6e4$ Bytes.  It is considered an Urban Macro (UMa) channel model with all the nodes in non-line-of-sight and one bandwidth part with a central frequency of $4$ GHz and bandwidth of $20$ MHZ. In the next subsection, we will discuss the performance results of the proposed scheme.  

\begin{table}
\caption{Reinforcement Learning Settings.}

\begin{center}
\resizebox{\columnwidth}{!}{%
\begin{tabular}{c c} 
\hline
\textbf{Parameter}&\textbf{Value} \\
\hline

Maximum training steps & {$30\mathrm{e}{4}$} \\
Initial $\epsilon$ & {1} \\
$\epsilon$ decay steps & {$15e3$ steps} \\
$\epsilon$ minimum & {$5\mathrm{e}{-2}$} \\
Buffer $(\mathcal{D})$ size & {$2\mathrm{e}{3}$} \\
Batch size & {$64$} \\ 
Learning rate & {$8\mathrm{e}{-3}$} \\
Discount factor $(\gamma)$ & {$0.99$} \\
Recurrent Layer hidden dimension & {$64$} \\
MultiLayer Perceptron hidden dimension & {$64$} \\
Weight initializer & { Orthogonal } \\
Deep Q-Network structure & {Double Q-networks} \\
Optimistic Weight $(\alpha)$ & {$0.1$} \\
The number of layers of hyper-network  & {2} \\
Hidden layer of the hyper-network  & {64} \\
\hline
Number of agents in the MAS & {$3$}\\
Transmission RLC buffer $(\bm{b})$ & {$\{0-0.96,0.96-0.97,0.97-98$}\\
occupancy ratio ranges  & {$0.98-0.99,0.99-1\}$}\\
Observation window ($t_w$) & {$0.5$ s}\\
\hline
\end{tabular}

}

\label{learning_settings}
\end{center}
\vspace{-5mm}
\end{table}

\begin{table}
\caption{Network Settings}
\begin{center}
\resizebox{\columnwidth}{!}{%

\begin{tabular}{c c} 
\hline
\textbf{Parameter}&\textbf{Value} \\
\hline
Wireless network & { New Radio (NR) } \\
Channel Bandwidth & { $40$ MHz } \\
Central Frequency ($f_c$) & { $4$ GHz } \\
Number of UEs $(N)$ & { 3 } \\
Number of gNB & {$1$}\\
Propagation Loss Model & { UMa nLos }\\
Numerology & { 2 } \\
gNB Noise Figure & {$5$ dB}\\
gNB Transmission Power & {$43$ dBm}\\
gnB Antenna configuration & {4x8} \\
UE Noise Figure & {$7$ dB}\\
UE Transmission Power & {$26$ dBm}\\
UE Antenna configuration & {1x1} \\
Max Transmission Buffer Size & {60 KBytes} \\
XR and CG traffic characteristics & {AR (3 flows), VR (1 flow), CG (1 flow) } \\
Codec data rate [min,max] & {AR: [0.5,10], VR: [10,30], CG: [10,30] Mbps } \\
\hline
\end{tabular}
}
\label{net_settings}

\end{center}
\vspace{-7mm}
\end{table}

\subsection{Simulation Results}
We present the performance results of our proposed schemes in terms of RL convergence, $\%$ of success, and KPIs as XR index, jitter, delay, and PLR versus distance. For each distance, the data of 10 runs is collected and a $95\%$ confidence interval is considered. In addition, we show flow-based graphs of the throughput and goodput. It is worth mentioning, that the proposed algorithm in \cite{Bojovic2023},  APS performs the best among all schemes in that work. Interestingly, the results obtained in such previous research show that when channel conditions are bad, the algorithm sends bigger frames. This behavior is undesired and must be avoided since increasing the XR packet size when the channel conditions are not favorable due to low Signal-to-Interference-Noise Ratio (SINR) is not the best strategy. Figure \ref{convergence} shows the convergence and $\%$ of success for three different distances for algorithms QMIX and oQMIX, respectively. The success rate is defined as the percentage of completion of the simulation without triggering the done condition as in algorithm \ref{oqmix}. As expected, when UEs are closer to the gNB as in Fig. \ref{convergence} (a) and (d), the success rate is almost maximum with a short number of episodes. Meaning, that solving the problem when the signal-to-interference and noise ratio (SINR) is low becomes easy for our scheme.

On the other hand, when the distance increases, the problem solution complexity also grows. This is related to the fact that only a small set of actions will be the ones that satisfy the requirements imposed by the reward function when stringent requirements are set. In Fig. \ref{convergence} (b,e) and (c,f) can be observed that the reward convergence time and success rate increase considerably for both schemes. However, note that oQMIX offers a faster convergence than QMIX. This can be easily spotted by observing the number of episodes where convergence is achieved. For instance, in Fig. \ref{convergence} (e) and (b) for reference episode 500 the value achieved in reward and percentage of success is lower for QMIX. The same can be seen if we look at Fig. \ref{convergence} (f) and (c) for the reference episode 1000.

Furthermore, in Fig. \ref{xr_jitt_del_plr_} we present the results in terms of  XR index, jitter, delay, and PLR. We observe that when UEs are in the $200$ and $300$ m range, APS, QMIX, and oQMIX perform similarly in terms of the XR index with a small gain of $1.7 \%$ of oQMIX algorithm. Conversely, oQMIX and QMIX offer an improvement when the distance increases in terms of average XR index up to $30.1 \%$ and $17.6\%$, respectively. For other KPIs of interest, oQMIX and QMIX provide average gains over APS of $15.6 \%$,  $16.5 \%$ $50.3 \%$, and $13.2 \%$,  $11.2 \%$, $7.86 \%$ with respect jitter, delay, and PLR, respectively.

\begin{figure*}[t]
\center
  \includegraphics[scale=0.55]{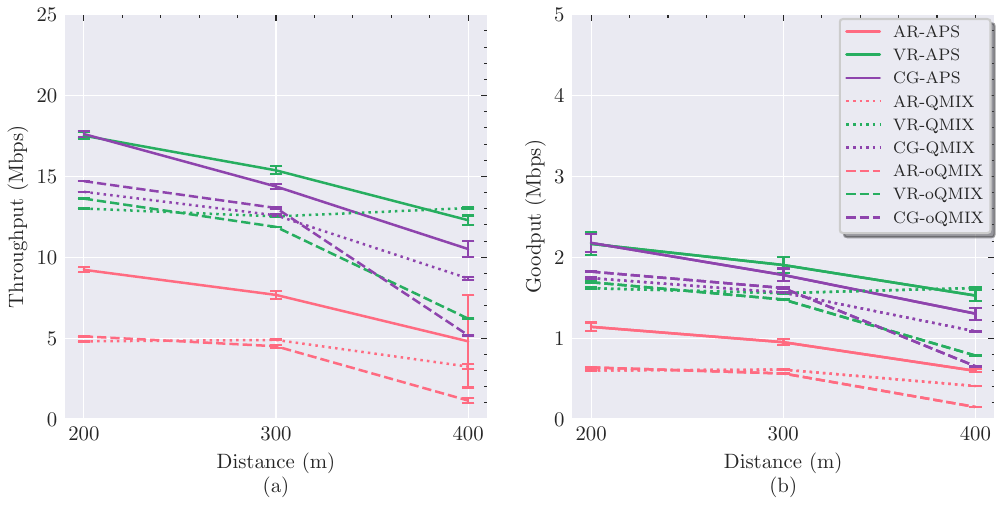}
  \setlength{\belowcaptionskip}{-5pt}
  \caption{Flow-based performance of APS, QMIX and oQMIX vs. Distance $\bm{(a)}$ Throughput and $\bm{(b)}$ Goodput } 
  \label{flow_graphs}
  \vspace{0mm}
\end{figure*}

In Fig. \ref{flow_graphs} we show a comparison of throughput and goodput per XR flow versus distance. In Fig. \ref{flow_graphs} (a), it is evident that the baseline APS exhibits a more aggressive behavior in data rate selection as throughput increases, while oQMIX and QMIX demonstrate a more conservative decision-making approach. Furthermore, Fig. \ref{flow_graphs} (b) illustrates the performance in terms of goodput for both the baseline APS and the proposed RL schemes. It's important to note that goodput is calculated at the application layer, whereas throughput is calculated at the transport layer. Although APS shows slightly better goodput performance, this is primarily attributed to its aggressive behavior in increasing data rates, which leads to performance degradation in PLR as distance increases. The effects of increasing data rate and the consequent growth of PLR, delay, and jitter become more pronounced as the number of XR users increases. This is particularly undesirable in XR applications where maintaining low delay and jitter are crucial for ensuring a high Quality of Experience (QoE) for the end user. The results without the attention mechanism exhibited poor performance, including a lack of convergence and inadequate KPI performance. Therefore, they were not included in any of the presented figures.

\section{Conclusions } \label{Section6}

In this paper, we presented a Multi-Agent Reinforcement Learning (MARL) algorithm with Attention Action Selection to improve Extended Reality (XR) Key Performance Indicators (KPIs) of interest. More specifically, the proposed algorithm named Optimistic QMIX (oQMIX) uses attention action selection to reduce the action set comprised by the codec parameters of the different new generation traffic types: Augmented Reality (AR), Virtual Reality (VR), and Cloud Gaming (CG). We presented the results in terms of convergence and $\%$ of success of three different scenarios with varying distances. Results showed that when the distance between the UE and the base station increases, the problem difficulty grows and convergence time also presents the same behavior.  Furthermore, we presented the simulation results of our proposed scheme and a state-of-the-art baseline Adjust Packet Size (APS). Results show that oQMIX overperforms APS with an average gain of $30.1.\%$, $15.6 \%$, $16.5 \%$ $50.3 \%$ for XR index, jitter, delay, and PLR, respectively.
On the other hand, we observed that APS presented a more aggressive behavior with the tendency of higher throughput in all XR and CG flows, increasing packet collisions and packet loss when distance increased. Conversely, oQMIX presented a more conservative behavior reducing PLR and maintaining a similar behavior in terms of goodput for both under-study algorithms. In future work, we intend to improve the attention algorithm and extend our study with numerous XR UEs.  

\section{Acknowledgment }\label{Section7}
This research is supported by the NSERC Canada Research Chairs program, Mitacs Accelerate Program, and Ericsson.\vspace{-2mm}

%---

%----------------------------------------------------------------------------------------
%	BIBLIOGRAPHY
%----------------------------------------------------------------------------------------

\bibliography{biblio.bib}{}
\bibliographystyle{IEEEtran}
% \printbibliography[title={Bibliography}] % Print the

\end{document}